\documentclass[aps,pra,amsmath,amssymb,reprint,floatfix]{revtex4-1}

\usepackage{xcolor}
\usepackage[]{graphicx}
\graphicspath{{./images/}}
\usepackage{xspace}
\usepackage{dsfont}
\usepackage{hyperref}

\newcommand{\ket}[1]{\vert #1 \rangle} 
\newcommand{\bra}[1]{\langle #1 \vert} 
\newcommand{\braket}[2]{\langle #1 | #2 \rangle} 
\newcommand{\ketbra}[2]{\left| #1 \middle> \! \middle< #2
\right|} 
\newcommand{\matrixel}[3]{\langle #1 \vphantom{#3} | #2 | #3
\vphantom{#1} \rangle} 
\newcommand{\abs}[1]{\vert #1 \vert} 
\newcommand{\I}{\mathrm{i}} 
\newcommand{\unit}[1]{\ensuremath{\, \mathrm{#1}}}
\newcommand{\commutator}[2]{[ #1 , #2 ]} 
\newcommand{\br}[1]{\left( #1 \right)}

\newcommand{\brrr}[1]{\left\{ #1 \right\}}

\newcommand{\de}{\mathrm{d}}

\DeclareMathOperator{\Tr}{Tr}
\DeclareMathOperator{\sech}{sech}

\newcommand{\Lio}{\mathcal{L}} 
\newcommand{\Id}{\mathds{1}}

\newcommand{\Ein}{\mathcal{E}_\mathrm{in}}

\newcommand{\E}{\mathcal{E}}

\newcommand{\J}{\mathcal{J}}
\newcommand{\Tc}{T_\mathrm{c}}
\newcommand{\MHz}{\times2\pi\unit{MHz}}
\newcommand{\wc}{\omega_\mathrm{c}}
\newcommand{\wl}{\omega_\mathrm{L}}
\newcommand{\kbad}{{\kappa_\mathrm{loss}}}
\newcommand{\ktot}{{\kappa_\mathrm{tot}}}

\begin{document}

\title{Weak coherent pulses for single-photon quantum memories}

\author{Luigi Giannelli}

\author{Tom Schmit}

\author{Giovanna Morigi} \affiliation{Theoretische Physik, Universit\"at des
  Saarlandes, 66123 Saarbr\"ucken, Germany}

\date{\today}

\begin{abstract}
  Attenuated laser pulses are often employed in place for single photons in
  order to test the efficiency of the elements of a quantum network. In this
  work we analyse theoretically the dynamics of storage of an attenuated light
  pulse (where the pulse intensity is at the single photon level) propagating
  along a transmission line and impinging on the mirror of a high finesse
  cavity. Storage is realised by the controlled transfer of the photonic
  excitations into a metastable state of an atom confined inside the cavity and
  occurs via a Raman transition with a suitably tailored laser pulse, which
  drives the atom and minimizes reflection at the cavity mirror. We determine
  the storage efficiency of the weak coherent pulse which is reached by
  protocols optimized for single-photon storage. We determine the figures of
  merit and we identify the conditions on an arbitrary pulse for which the
  storage dynamics approaches the one of a single photon. Our formalism can be
  extended to arbitrary types of input pulses and to quantum memories composed
  by spin ensembles, and serves as a basis for identifying the optimal protocols
  for storage and readout.
\end{abstract}


\maketitle

\section{\label{sec:introduction}Introduction}
Single photons are important elements for secure communication using light
\cite{Afzelius2015,Sangouard2012}. Integrating single photons in a quantum
network~\cite{Ritter2012}, on the other hand, requires stable and efficient
single photon sources, reliable storage units such as single-photon quantum
memories, quantum information processors, and ideally dissipationless
transmission channels~\cite{Briegel1998,Kurizki2015}. Since these devices
usually optimally work in different frequency regimes, the realization of
efficient quantum networks implies the ability of interfacing hybrid elements
\cite{Kurizki2015,Uphoff2016}. Proof-of-principle experiments for quantum
memories have therefore often made use of pulses generated by stable lasers at
the required frequency
\cite{Choi2008,Kimble2008,Usmani2010,Lan2009,Gisin2007,Koerber2018}. The laser
pulses are typically attenuated to the regime where the probability that they
contain a single photon is very small, while the probability that two or more
photons are detected is practically negligible. Even though photo-detection
after a beam splitter shows the granular properties of the light, yet the
coherence properties of weak laser pulses are quite different from the ones of
a single photon~\cite{Mack2003}. In particular, they are well described by
coherent states of the electromagnetic field, whose correlation functions can
be reproduced by a classical coherent field
\cite{Glauber1963,Glauber1963a,Schleich2001}. In this perspective it is
therefore legitimate to ask which specific information about the efficiency of
a single-photon quantum network can one possibly extract by means of weak laser
pulses.

Theoretically, similar questions have been analysed in
Ref.~\cite{Fleischhauer2000,Gorshkov2007,Gorshkov2007a,Gorshkov2007b,Dilley2012,Kalachev2007,Kalachev2008,Kalachev2010}.
In~\cite{Fleischhauer2000,Gorshkov2007,Gorshkov2007a,Gorshkov2007b,Dilley2012},
in particular, the authors consider a quantum memory composed by an atomic
ensemble, where the number of atoms is much larger than the mean number of
photons of the incident pulse. In this limit the equations describing the
dynamics can be brought to the form of the equations describing the interaction
of a single photon with the medium, and one can simply extract from the study
of one case the efficiency of the other. This scenario changes dramatically if
the memory is composed by a single
atom~\cite{Cirac1997,Reiserer2015,Duan2010,Kurz2014}. In this case the dynamics
is quite different depending on whether the atom interacts with a single photon
or with (the superposition of) several photonic excitations.

\begin{figure}[!ht]
  \centering \includegraphics[width=0.83\linewidth]{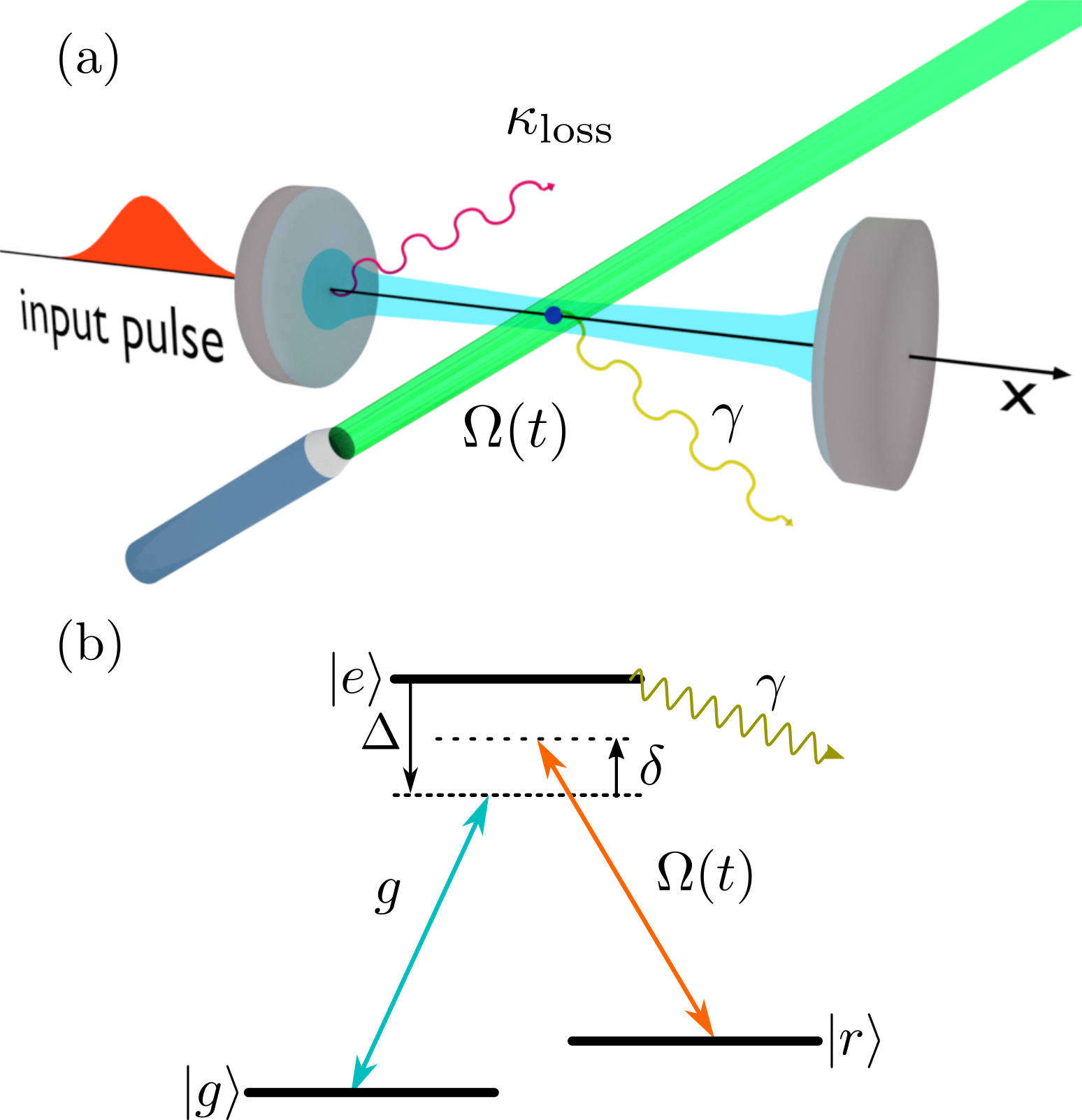}
  \caption{\label{Fig:1}(Color online) An input pulse propagates along a
    transmission line and impinges onto a cavity mirror (a). The pulse is
    absorbed and the atom undergoes a Raman transition from the initial state
    $|g\rangle$ to the final state $|r\rangle$ via the common excited state
    $|e\rangle$. This dynamics occurs thanks to a laser with appropriately
    tailored Rabi frequency $\Omega(t)$, which drives the transition
    $\ket{r}\to\ket{e}$ in order to maximize the transfer to state $\ket r$ and
    simultaneously minimize reflection at the mirror. We analyse the dynamics
    of storage when the incident light is described by a weak coherent pulse
    and $\Omega(t)$ is designed to optimize storage of a single photon. Further
    parameters are defined in the text.}
\end{figure}

In this work we theoretically analyse the dynamics of the storage of a
\textit{weak coherent pulse} into the excitation of a single-atom confined
within an optical resonator like in the setups
of~\cite{Specht2011,Khudaverdyan2008,Kimble1998,Keller2004}. The laser pulse
propagates along a transmission line and impinges on the mirror of the
resonator, as illustrated in Fig. \ref{Fig:1}(a). A control laser drives the
atom in order to optimize the transfer of the propagating pulse into the atomic
excitation $\ket{r}$, as shown in Fig. \ref{Fig:1}(b). We determine the
efficiency of storage under the assumption that the control laser optimizes the
storage of a single photon, which possesses the same time dependent amplitude
as the weak coherent pulse. Our goal is to identify the regime and the
conditions for which the dynamics of storage of the weak coherent pulse
reproduces the one of a single photon. This study draws on the protocols based
on adiabatic transfer identified in
Refs.~\cite{Fleischhauer2000,Gorshkov2007a,Dilley2012,Giannelli2018}. The
theoretical formalism for the interface between the weak coherent pulse
propagating along the transmission line and the single atom inside the
resonator is quite general and can be extended to describe the storage fidelity
of an arbitrary quantum state of light into excitations of the memory.

This manuscript is organized as follows. In Sec.~\ref{sec:model} we introduce
the theoretical model. In Sec.~\ref{sec:storage} we report our results: in
Sec~\ref{sec:numerical-results} we analyse the storage fidelity of a weak
coherent pulse. In Sec.~\ref{sec:effect-hamilt} we analyze the storage fidelity
of an arbitrary incident pulse at the single photon level. We then compare them
with the storage fidelity of a single photon. The conclusions are drawn in
Sec.~\ref{Sec:Conclusions}. The appendices provide details to the calculations
in Secs.~\ref{sec:model} and~\ref{sec:storage}.

\section{\label{sec:model}Basic model}
Figure~\ref{Fig:1} reports the basic elements of the dynamics. A weak coherent
pulse propagates along the transmission line and impinges on the mirror of a
optical high-finesse cavity. Here it is transmitted into a cavity mode at
frequency $\wc$, which, in turn, interacts with a single atom confined within
the resonator. The atom is driven by a laser, whose temporal shape is tailored
in order to maximize the transfer of a single photonic excitation, with the
same amplitude as the weak coherent pulse, into an atomic excitation $\ket{r}$.

In the following we provide the details of the theoretical model and we
introduce the physical quantities which are important for the discussion of the
rest of this paper.

\subsection{Master equation\label{sec:hamiltonian}}
We describe the dynamics of storage by determining the density matrix
$\hat{\rho}$ for the cavity mode, the atom, and the modes of the transmission
line. Its evolution is governed by the master equation ($\hbar=1$)
\begin{equation}
  \label{eq:mastereq}
  \partial_t\hat{\rho} = -\I\commutator{\hat{H}_\mathrm{tot}(t)}{\hat{\rho}}
  +\Lio_\mathrm{dis}\hat{\rho}\,,
\end{equation}
where Hamiltonian $\hat{H}_\mathrm{tot}(t)$ determines the coherent evolution
and superoperator $\Lio_\mathrm{dis}$ the incoherent dynamics. Below we define
them.

The Hamiltonian $\hat H$ describes the unitary dynamics of the system composed
of the modes of the transmission line, the cavity mode, and the atom's internal
degrees of freedom. We decompose it into the sum of two terms
\begin{equation}
  \label{eq:h}
  \hat{H}_\mathrm{tot}(t)= \hat{H}_\mathrm{fields}+\hat{H}_\mathrm{I}(t)\,.
\end{equation}
The term $\hat{H}_\mathrm{fields}$ describes the coherent dynamics of the
electromagnetic fields in absence of the atom. In the reference frame rotating
at the cavity mode frequency $\wc$ it reads
\begin{equation}
  \label{eq:hf}
  \hat{H}_\mathrm{fields}=\sum_k(\omega_k-\wc)\hat{b}_k^\dag \hat{b}_k+
  \sum_k\lambda_k(\hat{a}^\dag \hat{b}_k+\hat{b}_k^\dag \hat{a}).
\end{equation}
Here, $\omega_k$ are the frequencies of the electromagnetic field's modes of
the transmission line, operators $\hat{b}_k$ and $\hat{b}_k^\dag$ annihilate
and create, respectively, a photon at frequency $\omega_k$, with
$\commutator{\hat{b}_k}{\hat{b}_{k'}^\dag} = \delta_{k,k'}$. The modes
$\hat{b}_k$ are formally obtained by quantizing the electromagnetic field in a
resonator of length $L$, where $L$ is taken to be much larger than any other
length in the system. They are standing wave modes with a node at the cavity
mirror (here at $x=0$) and have the same polarization as the cavity mode (see
Appendix~\ref{sec:descr-electr-field}). The latter is described by a harmonic
oscillator with annihilation and creation operators $a$ and $a^\dag$, where
$\commutator{\hat{a}}{\hat{a}^\dag} = 1$ and $\commutator{\hat{a}}{\hat{b}_k} =
\commutator{\hat{a}}{\hat{b}_k^\dag} = 0$. In the rotating-wave approximation
the interaction is of beam-splitter type and conserves the total number of
excitations. The couplings $\lambda_k$ are related to the radiative damping
rate $\kappa$ of the cavity mode by $\kappa = L|\lambda(\wc)|^2/c$, with
$\lambda(\wc)$ the coupling strength at the cavity-mode resonance
frequency~\cite{Carmichael}. Furthermore, using the Markov approximation, the
couplings are taken to be $\lambda_k=\lambda(\wc)$.

The atom-photon interactions are treated in the dipole and rotating-wave
approximations. The fields interact with two dipolar transitions sharing the
common excited state $\ket e$, forming a $\Lambda$ level scheme, see Fig.
\ref{Fig:1}(b). The transition $|g\rangle\to |e\rangle$ couples with the cavity
mode with strength (vacuum Rabi frequency) $g$. Transition $|r\rangle\to
|e\rangle$ is driven by a laser with the time-dependent Rabi frequency
$\Omega(t)$. The corresponding Hamiltonian reads
\begin{equation}
  \label{eq:hat}
  \begin{aligned}
    \hat{H}_\mathrm{I}=\delta\ketbra{r}{r}-\Delta\ketbra{e}{e}+
    \left[g\ketbra{e}{g}\hat{a}+\Omega(t)\ketbra{e}{r}+\mathrm{H.c.}\right],
  \end{aligned}
\end{equation}
where $\Delta = \wc - \omega_e$ is the detuning between the cavity frequency
$\wc$ and the frequency $\omega_e$ of the $\ket{g}-\ket{e}$ transition, while
$\delta = \omega_r+\wl-\wc$ is the two-photon detuning which is evaluated using
the central frequency $\wl$ of the driving field $\Omega(t)$. Here, $\omega_r$
denotes the frequency difference (Bohr frequency) between the state $\ket{r}$
and the state $\ket{g}$. Unless otherwise stated, in the following we assume
that the conditions of one and two-photon resonance $\Delta=\delta=0$ are
fulfilled.

Superoperator $\Lio_\mathrm{dis}$ describes the incoherent dynamics due to
spontaneous decay of the atomic excited state $\ket{e}$ at rate $\gamma$, and
due to the finite transmittivity of the second cavity mirror as well as due to
scattering and/or finite absorption of radiation at the mirror surfaces at rate
$\kbad$. We model each of these phenomena by Born-Markov processes described by
the superoperators $\Lio_\gamma$ and $\Lio_{\kbad}$, respectively, such that $
\Lio_\mathrm{dis} = \Lio_\gamma + \Lio_{\kbad}$ and
\begin{subequations}\label{eq:liouvillian}
  \begin{gather}
    \label{eq:liouvilliangamma}
    \mathcal{L}_\gamma\hat{\rho} = \gamma(2\ketbra{\xi_e}{e}\hat{\rho}
    \ketbra{e}{\xi_e}-\ketbra{e}{e}\hat{\rho}-\hat{\rho}\ketbra{e}{e})\,, \\
    \label{eq:liouvilliankappabad}
    \mathcal{L}_\kbad\hat{\rho} = \kbad(2\hat{a}\hat{\rho} \hat{a}^\dag -
    \hat{a}^\dag \hat{a}\hat{\rho}-\hat{\rho} \hat{a}^\dag \hat{a})\,.
  \end{gather}
\end{subequations}
Here, $\ket{\xi_e}$ is an atomic state into which the excited state $\ket{e}$
decays, which is assumed to be different from $\ket{g}$ and $\ket{r}$.

\subsection{\label{sec:state-description}Initial state}
The total state of the system $\ket{\psi_t}$ at the initial time $t=t_1$ is
given by a weak coherent pulse in the transmission line, the empty optical
cavity, and the atom in state $\ket{g}$:
\begin{equation}
  \label{eq:psi0}
  \ket{\psi_{t_1}}=\ket{g}\otimes\ket{0}_c\otimes\ket{\psi^\mathrm{coh}},
\end{equation}
where $\ket{0}_c$ is the Fock state of the resonator with zero photons.

Below we specify in detail the state of the field. The incident light pulse is
characterized by the time-dependent operator $\hat D$, such that its state at
the interface with the optical resonator reads
\begin{equation}
  \label{eq:WCP_N_def}
  \ket{\psi^\mathrm{coh}}=\hat D(\{\alpha_k\})\ket{\mathrm{vac}}
\end{equation}
and $\ket{\mathrm{vac}}$ is the vacuum state of the external electromagnetic
field. Operator $\hat D(\{\alpha_k\})$ takes the form
\begin{equation}
  \label{eq:D}
  \hat D(\{\alpha_k\}) =
  \otimes_k\exp(\alpha_k \hat{b}_k^\dagger-\alpha_k^*\hat{b}_k)\,,
\end{equation}
where $\alpha_k$ is a complex scalar and the index $k$ runs over all modes of
the electromagnetic field with the same polarization. It thus generates a
multi-mode coherent state, whose mean photon number $n$ is
\begin{equation}
  \label{eq:alpha}
  n=\matrixel{\psi^\mathrm{coh}}{\sum_{k}\hat{b}_k^\dag \hat{b}_k}
  {\psi^\mathrm{coh}} =
  \sum_{k}\abs{\alpha_k}^2 \,.
\end{equation}
In the following we assume that $n\ll 1$, which is fulfilled when
$|\alpha_k|^2\ll1$ for all $k$. We will denote this a \textit{weak coherent
  pulse}. This state approximates a single-photon state since at first order in
$n$ it can be approximated by the expression
\begin{equation}
  \ket{\psi^\mathrm{coh}}\approx(1-n/2)\ket{\mathrm{vac}}+
  \sum_{k}\alpha_k\hat b_k^\dagger\ket{\mathrm{vac}}\,.
\end{equation}
Coefficients $\alpha_k$ are related to the pulse envelope $\Ein(t)$ at position
$x=0$ (which is the position of the mirror interfacing the cavity with the
transmission line) via the relation
\begin{equation}
  \label{eq:alpha_k}
  \alpha_k=\sqrt{\frac{c}{2L}}\int_{-\infty}^{\infty}\de t e^{\I(kc -\wc)t}
  \Ein(t)
\end{equation}
with $c$ the speed of light and $L$ the length of the transmission line. The
squared norm of $\Ein(t)$ equals the number of impinging photons in
Eq.~(\ref{eq:alpha}):
\begin{equation}
  \label{eq:Einnorm}
  \int_{-\infty}^\infty\abs{\Ein(t)}^2\de t =n\,.
\end{equation}
In this work we are interested in determining the storage efficiency of a weak
coherent pulse by the atom. We compare in particular the storage efficiency
with the one of a single photon, whose amplitude is given by the same amplitude
$\Ein(t)$, apart for a normalization factor giving that the integral in Eq.
\eqref{eq:Einnorm} is unity. For this specific study we choose
\begin{equation}
  \label{eq:Einhypsec}
  \Ein(t)=\frac{\sqrt{n}}{\sqrt{T}}\sech{\br{\frac{2t}{T}}}\,,
\end{equation}
where $T$ is the characteristic time determining the coherence time $\Tc=\pi
T/4\sqrt{3}$ of the light pulse, defined as
\begin{equation}
  \label{eq:Tc} 
  \Tc = \sqrt{\langle t^2\rangle - \langle t \rangle^2} 
\end{equation}
with $\langle t^x \rangle \equiv \int_{t_1}^{t_2}t^x\abs{\Ein(t)}^2\de t$. The
dynamics is analysed in the interval $t\in[t_1,t_2]$, with $t_1<0<t_2$ and
$|t_1|, t_2\gg \Tc$, such that (i) at the initial time there is no spatial
overlap between the input light pulse and the cavity mirror and (ii) at $t=t_2$
the reflected component of the light pulse is sufficiently far away from the
mirror so that it has no spatial overlap with the cavity mode. The choice of
these parameters has been discussed in detail in
Appendix~\ref{sec:descr-electr-field} and in Ref.~\cite{Giannelli2018}.

\subsection{Target dynamics}
\label{Sec:Target}
The target of the dynamics is to absorb a single photon excitation and populate
the atomic state $\ket{r}$. This dynamics is achieved by suitably tailoring the
control field $\Omega(t)$. We will consider protocols using control fields
$\Omega(t)$ that have been developed for a single-photon wave
packet~\cite{Fleischhauer2000,Gorshkov2007a,Dilley2012,Giannelli2018}. The
figures of merit we take are (i) the probability $\eta$ to find the excitation
in the state $\ket{r}$ of the atom after a fixed interaction time and (ii) the
fidelity of the transfer $\nu$, which we define as the ratio between the
probability $\eta$ and the number of impinging photons. This ratio, as we show
in the next section, approaches the fidelity of storage of a single photon
$\eta^\mathrm{sp}$ when $n\to 0$.

We give the formal definition of these two quantities. The probability $\eta$
reads \cite{Gorshkov2007a}
\begin{equation}
  \label{eq:eta}
  \eta = \Tr\brrr{\hat{\rho}(t_2)\ketbra{r}{r}\otimes{\Id_\mathrm{em}}}
  = \matrixel{r}{\Tr_\mathrm{em}\brrr{\hat{\rho}(t_2)}}{r}
\end{equation}
where $\Id_\mathrm{em}$ and $\Tr_\mathrm{em}$ denote respectively the identity
and the trace over the electromagnetic fields (both the fields in the
transmission line and in the optical cavity), and $\hat{\rho}(t)$ is the
density operator of the system.

The fidelity of the transfer is defined as the ratio between $\eta$ and the
number of impinging photons, namely
\begin{equation}
  \label{eq:nu}
  \nu=\frac{\eta}{\int_{t_1}^{t_2}\abs{\Ein(t)}^2\de t},
\end{equation}
which is strictly valid for a coherent pulse. This definition of the
  fidelity quantitatively describes the probability that the incident pulse is
  stored by the atom. It agrees with the definition of
  Ref.~\cite{Gorshkov2007a}, where the authors denote this quantity by
  ``efficiency''. Indeed, if the initial state is a single photon, the fidelity
  $\nu$ and the efficiency $\eta$ coincide.

Before we conclude, we remind the reader of the cooperativity $C$, which
determines the maximum fidelity of single-photon
storage~\cite{Gorshkov2007a,Giannelli2018}. The cooperativity $C$ characterizes
the strength of the coupling between the cavity mode and the atomic transition,
it reads~\cite{Gorshkov2007a}
\begin{equation}
  \label{eq:C}
  C = \frac{g^2}{\ktot\gamma}, 
\end{equation}
where $\ktot=\kappa+\kbad$ is the total cavity decay rate. For protocols
based on adiabatic transfer of the single photon into the atomic excitation,
the maximum fidelity of single-photon storage
reads~\cite{Gorshkov2007a,Giannelli2018}
\begin{equation}
  \label{eq:etasinglephoton}
  \eta_\mathrm{max}^\mathrm{sp}=\frac{\kappa}{\ktot}\frac{C}{1+C}\,,
\end{equation}
and it approaches $\kappa/\ktot$ for $C\to\infty$.
Equation~(\ref{eq:etasinglephoton}) is also the probability for emission of a
photon into the transmission line when the atom is initially prepared in the
excited state $\ket{e}$ and no control pulse is applied.

The parameters we use in our study are the ones of the setup of
Ref.~\cite{Koerber2018}, $(g,\kappa,\gamma,\kbad)=(4.9,2.42,3.03,0.33)\MHz$,
corresponding to the cooperativity $C\approx2.88$ and to the maximal storage
fidelity $\eta_\mathrm{max}^\mathrm{sp}\approx0.65$. Furthermore we choose
$\Tc=0.5\unit{\mu s}$ such that the adiabatic condition is fulfilled:
$\gamma\Tc C \approx 27 \gg1$ (see Ref.~\cite{Giannelli2018}).

\section{\label{sec:storage}Storage}
In this section we report the results of the storage of weak coherent pulses
into a single atom excitation. We first determine efficiency and fidelity by
numerically solving the master equation of Eq.~\eqref{eq:mastereq}. We compare
the results with the corresponding storage fidelity of a single photon with
temporal envelope $\Ein(t)$, Eq. \eqref{eq:Einhypsec}. We then determine
analytically the efficiency $\eta$ and the fidelity $\nu$ for weak coherent
pulses with mean photon number $n\ll 1$ and quantify the discrepancy between
these quantities and the single-photon storage fidelity as a function of $n$.
We further discuss how this method can be extended in order to determine the
efficiency of storage of an arbitrary incident pulse.

\subsection{\label{sec:numerical-results}Numerical results}
We determine the dynamics of storage by numerically integrating a master
equation in the reduced Hilbert space of cavity mode and atomic degrees of
freedom, which we obtain from the master equation \eqref{eq:mastereq} after
moving to the reference frame which displaces the multimode coherent state to
the vacuum. The procedure extends to an input multi-mode coherent state an
established procedure for describing the interaction of a quantum system with
an oscillator in a coherent state, see for instance \cite{Cohen-Tannoudji1994}.
We apply the unitary transformation $\hat D'(t)=\hat D(\{\alpha_k(t)\})$, where
operator $\hat D$ is given in Eq.\eqref{eq:D} and the arguments are
$\alpha_k\to \alpha_k(t)=\alpha_ke^{-\I(\omega_k-\wc)t}$. In this reference
frame the initial state of the electromagnetic field is the vacuum, the full
density matrix is given by $\hat \rho'(t)=\hat D'(t)^\dag\rho(t)\hat D'(t)$ and
its coherent dynamics is governed by Hamiltonian
\begin{equation}
  \label{eq:h_wcp}
  \begin{aligned}
    \hat{H}'(t)={}&\hat{H}_\mathrm{tot}(t)+
    \sqrt{2\kappa}\br{\Ein(t)\hat{a}^\dag + \Ein^*(t)\hat{a}}.
  \end{aligned}
\end{equation}
Here $\Ein(t)$ carries the information about the initial state of the
electromagnetic field and it is related to the amplitudes $\alpha_k$ by the
following equation (consistently with Eq.~(\ref{eq:alpha_k}))
\begin{equation}
  \label{eq:Einvsalphas}
  \Ein(t) = \sqrt{\frac{Lc}{2\pi^2}}
  \int_{-\infty}^\infty\alpha(k+k_c) e^{-\I kct}\de k.
\end{equation}
By using the Born-Markov approximation one can now trace out the degrees of
freedom of the electromagnetic field outside the resonator. The Hilbert space
is then reduced to the cavity mode and atom's degrees of freedom, the density
matrix which describes the state of this system is
\begin{equation}
  \hat{\tau}(t)=\Tr_\mathrm{ff}{\hat{\rho}'(t)}\,,
\end{equation}
where $\Tr_\mathrm{ff}$ denotes the partial trace with respect to the degrees
of freedom of the external electromagnetic field. Its dynamics is governed by
the master equation
\begin{equation}
  \label{eq:mastereq_wcp}
  \partial_t\hat{\tau} =
  -\I\commutator{\hat{H}(t)}{\hat{\tau}}+\Lio_\mathrm{\gamma}\hat{\tau}
  +\Lio_\mathrm{\ktot}\hat{\tau}.
\end{equation}
and superoperators $\Lio_\gamma$ and $\Lio_\ktot$ are defined in
Eqs.~(\ref{eq:liouvillian}), where now the cavity field is damped at rate
$\ktot=\kappa+\kbad$ and $\kappa$ is the linewidth due to radiative decay of
the cavity mode by the finite transmittivity of the mirror at $x=0$. The
initial state is here described by the density operator
$\hat{\tau}(t_1)=\ketbra{g,0_c}{g,0_c}$, and the storage efficiency is
$\eta=\Tr\brrr{\hat{\tau}(t_2)\ketbra{r}{r}}$.

\begin{figure}[!ht]
  \centering \includegraphics[]{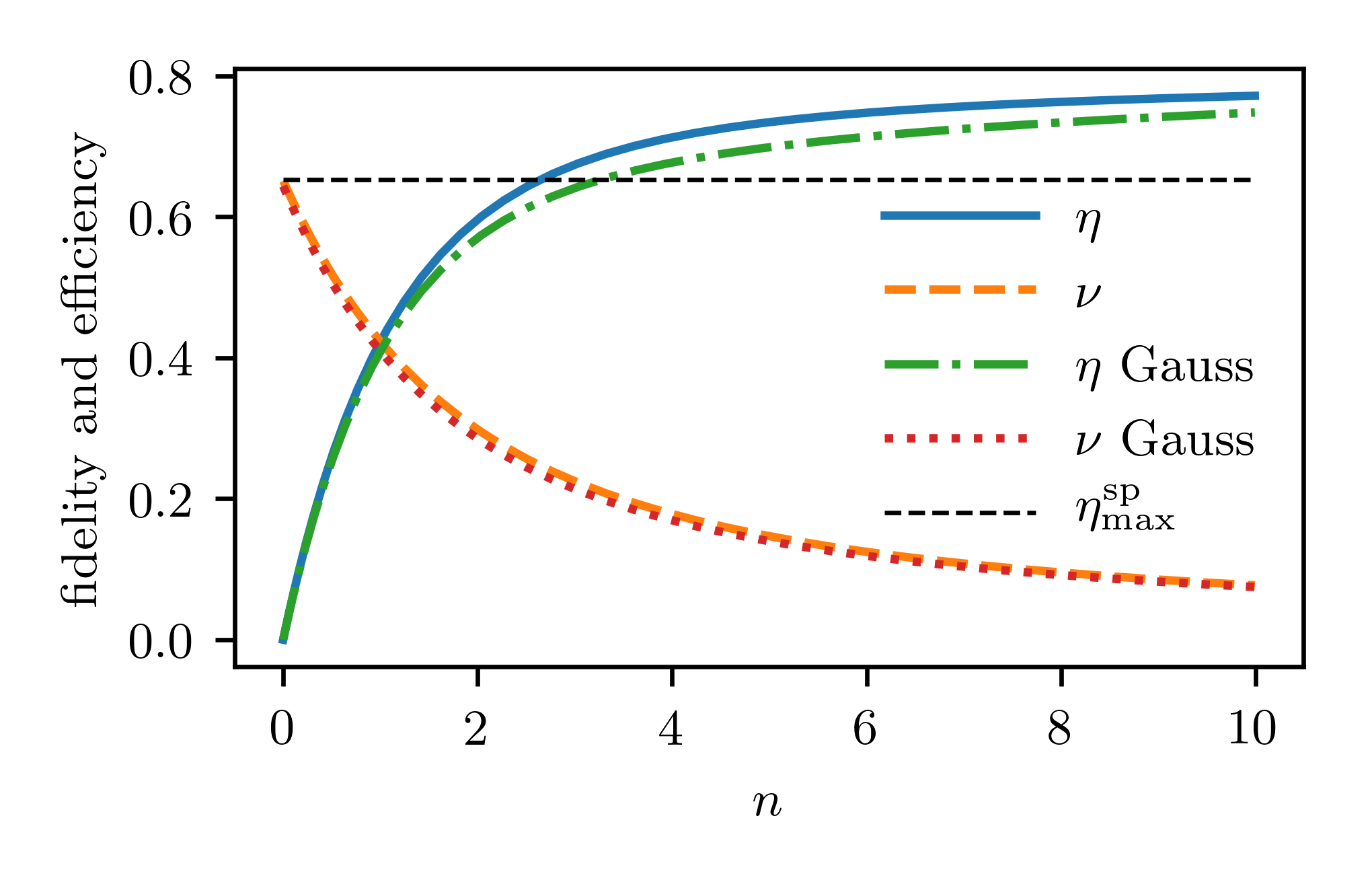}
  \caption{\label{fig:wcp_me}(Color online) Storage efficiency $\eta$, Eq.
    \eqref{eq:eta} and fidelity $\nu$, Eq. \eqref{eq:nu}, at time $t=t_2$ as a
    function of the mean photon number $n$ of the incident weak coherent pulse
    with shape of Eq.~(\ref{eq:Einhypsec}) (solid and dashed). The figures of
    merit $\eta$ and $\nu$ have been evaluated by determining numerically the
    density matrix of the system $\tau(t_2)$ from the initial state
    $\hat{\tau}(t_1)=\ketbra{g,0_c}{g,0_c}$ by integrating the master equation
    \eqref{eq:mastereq_wcp} in the truncated Hilbert space of the cavity field
    with a maximum of $14$ excitations. For comparison we also report the
    fidelity and efficiency of storage of a weak coherent pulse with Gaussian
    shape (labels ``Gauss''); In this case the control pulse is optimized for
    the storage of a single photon with Gaussian shape. The dashed line
    indicates the maximal fidelity of storage of a single photon. The other
    parameters are given in Sec. \ref{Sec:Target}.}
\end{figure}

We integrate numerically the optical Bloch Equation for the matrix elements of
Eq. \eqref{eq:mastereq_wcp} taking a truncated Hilbert space for the cavity
field, with number states ranging from $m=0$ to $m=m_{\rm max}$. For the
parameters we use in our simulation we find that the mean average number of
intracavity photons is below 2. We check the convergence of our simulation for
different values of $m=m_{\rm max}$ and fix $m_{\rm max}=14$. Figure
\ref{fig:wcp_me} displays the storage efficiency $\eta$ and fidelity $\nu$ at
time $t=t_2$ for different mean number of photons $n$ of the incident weak
coherent pulse. When evaluating the dynamics we employed the control laser
pulse $\Omega(t)$ which optimizes the storage of the incident pulse when this
is a {\it single} photon with temporal envelope $\Ein(t)$, Eq.
\eqref{eq:Einhypsec}. In detail, the amplitude of the laser pulse has been
determined in Ref.~\cite{Giannelli2018} and reads (for $\delta=\Delta=0$)
\begin{equation}
  \label{eq:OX}
  \Omega(t)=
  \sqrt{\frac{2\gamma(1+C)}
  {(e^{4t/T}+1)T}}.
\end{equation}
We observe that the storage efficiency $\eta$ rapidly increases with $n$ and
saturates to the asymptotic value $\eta_\infty\approx0.79$ for $n\gtrsim10$.
This asymptotic value indicates that the field in the cavity is essentially
classical, the dynamics is the one of STIRAP \cite{Vitanov2017}, and its
efficiency does not reach unity being the control pulse optimal for
single-photon storage but {\it not} for STIRAP. The fidelity $\nu$ decreases
with $n$, while in the limit $n\to 0$ it approaches the single-photon storage
fidelity. We note that the behavior for $n\gtrsim1$ depends on the pulse shape
(see Fig.~\ref{fig:wcp_me}).

In Ref.~\cite{Koerber2018} the authors report the experimental results of
measuring the fidelity $\nu$ as a function of $n$. In particular they report
the ratio between the fidelity of storing a weak coherent pulse with
$n\approx0.02$ and the fidelity for $n\approx1$ to be
$\nu_\mathrm{exp}(n=0.02)/\nu_\mathrm{exp}(n=1)\approx1.27$. We compare these
results with our predictions for $n\ll1$ where the fidelity is independent of
the photon shape. Then, we extract the same ratio from Fig.~\ref{fig:wcp_me}
and obtain $\nu(n=0.02)/\nu(n=1)\approx1.5$. Even if for $n=1$ the fidelity
depends on the pulse shape, we have verified by comparing with different pulse
shapes that the discrepancy is typically small.

\subsection{\label{sec:effect-hamilt}Extracting the single-photon storage
  fidelity from arbitrary incident pulses}
The method we applied in Sec.~\ref{sec:numerical-results} is convenient but
valid solely when the input pulse is a coherent state. We now show a more
general approach for describing storage of a generic input pulse by an atomic
medium (which can also be composed by a single atom) and which allows to obtain
a useful description of the dynamics. This approach does not make use of
approximations such as treating the atomic polarization as an
oscillator~\cite{Gorshkov2007a} and allows one to determine the storage
fidelity.

For this purpose we consider master equation~(\ref{eq:mastereq}), and recast it
in the form~\cite{Moelmer1988,Dum1992}
\begin{equation}
  \label{eq:mastereq_recasted}
  \partial_t\hat{\rho} = -\I(\hat{H}_\mathrm{eff}(t)\hat{\rho}
  -\hat{\rho}\hat{H}_\mathrm{eff}^\dag(t)) +\J\hat{\rho},
\end{equation}
where $ \hat{H}_\mathrm{eff}(t)$ is a non-Hermitian operator, which reads
\begin{equation}
  \label{eq:Heff}
  \hat{H}_\mathrm{eff}(t) = \hat{H}_\mathrm{tot}(t) -\I\gamma\ketbra{e}{e}
  -\I\kbad\hat{a}^\dag\hat{a}\,,
\end{equation}
and is denoted in the literature as effective Hamiltonian. The last term on the
right-hand side of Eq.~(\ref{eq:mastereq_recasted}) is denoted by \textit{jump
  term} and is here given by
\begin{equation}
  \label{eq:J}
  \J\hat{\rho} = 2\br{\gamma\ketbra{\xi_e}{e}\hat{\rho}
    \ketbra{e}{\xi_e} + \kbad\hat{a}\hat{\rho} \hat{a}^\dag}.
\end{equation}
This decomposition allows one to visualize the dynamics in terms of an ensemble
of trajectories contributing to the dynamics, where each trajectory is
characterized by a number of jumps at given instant of time within the interval
where the evolution occurs~\cite{Dum1992,Carmichael}. Of all trajectories, we
restrict to the one where no jump occurs since this is the only trajectory
which contributes to the target dynamics. In fact, even though
  trajectories with spontaneous emission events may lead to dynamics where the
  atom is finally in state $\ket{r}$, yet such trajectories are incoherent and
  thus irreversible. We therefore discard them since they do not contribute to
  the fidelity of the process. The corresponding density matrix is
$\rho_0(t)=S(t)\rho(t_1)S(t)^\dagger/P_0$, where
$S(t)=T:\exp\left(-i\int_{t_1}^t d\tau \hat{H}_\mathrm{eff}(\tau)/\hbar\right)$
and $T$ is the time ordering operator, while $P_0={\rm
  Tr}\{S(t)\rho(t_1)S(t)^\dagger\}$ is the probability that the trajectory
occurs. Since the initial state is a pure state, $\hat{\rho}(t_1)
=\ketbra{\psi_{t_1}}{\psi_{t_1}}$, then
$\hat{\rho}_0(t)=\ketbra{\psi_t}{\psi_t}$ with
$\ket{\psi_t}=S(t)\ket{\psi_{t_1}}/\sqrt{P_0}$. The efficiency of storage
$\eta$, in particular, can be written as
\begin{equation}
  \label{eq:eta:Heff}
  \eta=P_0{\rm Tr}\{|r\rangle\langle r|\rho_0(t_2)\}\,.
\end{equation}
We note that this definition can be extended also to input pulses which
  are described by mixed states. In fact, consider the density matrix $\mu$ of
  the incident pulse: $\mu=\sum_\alpha
  p_\alpha\ketbra{\psi^\alpha}{\psi^\alpha}$, with $\sum_\alpha p_\alpha=1$ and
  each $\ket{\psi^\alpha}$ a quantum state of the electromagnetic field. The
  efficiency of storage of the mixed state $\mu$ is then
  \begin{equation}
    \label{eq:fidelitymixed}
    \eta^{\mathrm{mix}} = \sum_\alpha p_\alpha\eta^\alpha\,.
  \end{equation}
  Here, $\eta^\alpha$ is the efficiency of storage of the pure state
  $\ket{\psi^\alpha}$ which can be computed using Eq.~(\ref{eq:eta:Heff}).

In order to determine $\eta$, we first decompose the incident pulse at $t=t_1$
into photonic excitations, namely:
\begin{equation}
  \label{eq:WCP_N_deco_app}
  \ket{\psi^{\rm coh}} =
  \sum_{m=0}^\infty C_m
  \ket{\psi^{(m)}},
\end{equation}
where $\sum_m|C_m|^2=1$, and the state $\ket{\psi^{(m)}}$ contains exactly $m$
photons, $\braket{\psi^{(\ell)}}{\psi^{(m)}}=\delta_{\ell,m}$. The dynamics
transfers the excitations but preserves their total number, since $
\hat{H}_\mathrm{eff}$ commutes with $\sum_kb_k^\dagger b_k+a^\dagger
a+\ketbra{e}{e}+\ketbra{r}{r}$. Therefore it does not couple states
$\ket{\psi^{(m)}}$ with different number of photons. By this decomposition we
can numerically determine the fidelity $\eta$ for a finite number of initial
excitations, as we show in Appendix~\ref{App:B}. The efficiency $\eta$, in
particular, can be cast in the form
\begin{equation}
  \label{eq:fidelityexpansion}
  \eta= \sum_{m=0}^\infty |C_m|^2\eta^{(m)}\,,
\end{equation}
where $\eta^{(m)}=\bra{\psi^{(m)}}S(t)^\dagger|r\rangle\langle
r|S(t)\ket{\psi^{(m)}}$ is the efficiency that one photon from a $m$-photon
state is transferred into the atomic excitation $\ket{r}$. Here, $\eta^{(1)}$
is the storage fidelity of a single photon $\eta^\mathrm{sp}$. For a weak
coherent pulse $C_m= \sqrt{e^{-n} n^{m}/m!}$, and for $n\ll 1$ we obtain the
expression
\begin{equation}
  \label{eq:pop2vsalpha}
  \eta = n \eta^{(1)} +n^2\left( \eta^{(2)}/2-\eta^{(1)}\right) + O(n^3)\,.
\end{equation}
such that the fidelity takes the form
\begin{equation}
  \label{eq:fid2vsalpha}
  \nu = \frac{\eta}{n} =  \eta^{(1)} +n\left( \eta^{(2)}/2-\eta^{(1)}\right) + O(n^2)\,.
\end{equation}
If the control pulse $\Omega(t)$ is chosen to be the one which maximize the
storage fidelity of a single photon, then
$\eta^{(1)}=\eta^\mathrm{sp}_\mathrm{max}$, Eq.~(\ref{eq:etasinglephoton}).
This can be clearly seen in Fig.~\ref{fig:wcp_me}.

We now discuss this dynamics if, instead of a single atom, the quantum memory
is composed by $M$ atoms within the resonator. In the following we assume that
the atoms are identical and that the vacuum Rabi coupling and the control laser
pulse intensity and phase do not depend on the atomic positions within the
cavity. Let us first consider that the input pulse is a single photon. In this
case the dynamics can be mapped to the one described by Eq.
\eqref{eq:mastereq}, where in the Hamiltonian \eqref{eq:hat} the states of the
$\Lambda$ transition are replaced by the collective atomic states $\ket{g}\to
\ket{g'}=\ket{g_1,\ldots,g_M}$, $\ket{e}\to
\ket{e'}=\sum_{i=1}^M\ket{g_1,\dots,e_i,\dots,g_M}/\sqrt{M}$, and
$\ket{r}\to\ket{r'} = \sum_{i=1}^M\ket{g_1,\dots,r_i,\dots,g_M}/\sqrt{M}$,
where the latter is the target state. For a single incident photon, in fact,
these are the only internal states involved in the dynamics. The coupling
between the cavity mode and the $\ket{g'}-\ket{e'}$ transition is now
$g\sqrt{M}$, leading to a higher cooperativity $C$ and thus to a larger value
of $\eta_\mathrm{max}^\mathrm{sp}$. In this case the control pulse leading to
optimal storage is the same as for a single atom, which couples to the cavity
with vacuum Rabi frequency $\tilde g=g\sqrt{M}$ (see for example
Eq.~(\ref{eq:OX}) and Ref.~\cite{Giannelli2018}).

If the incident pulse is not a single photon, further collective excitations of
the atoms have to be accounted for and the dynamics cannot be reduced to the
coupling of a $\Lambda$ structure with the cavity field, as is detailed in
Appendix \ref{App:B} for the case of a weak coherent pulse. Nevertheless, if
the number of atoms is much larger than the mean number of excitations in the
incident pulse $M\gg n$, the dynamical equations can be reduced to the ones
describing storage of the single
photon~\cite{Fleischhauer2000,Gorshkov2007a,Dilley2012}. In this limiting case,
the optimal control pulses for storage of a single photon can also be applied
to storage of the input pulse by the atomic ensemble, as long as the input
pulse has the same envelope as the single photon. We refer the interested
reader to Ref. \cite{Gorshkov2007a} for details.

In general, the formalism of the effective Hamiltonian can be applied to
determine the control field for storage of an arbitrary input pulse by an
atomic ensemble, without having to impose the condition $M\gg n$. For an
arbitrary input pulse, $\ket{\psi} = \sum_{m=0}^\infty C_m \ket{\psi^{(m)}}$
with $\sum_m|C_m|^2=1$, the target state is $\sum_{m=0}^\infty C_m \ket{r_m}$,
where $|r_m\rangle$ is the Dicke state of the atomic ensemble where $m$ atoms
are in $|r\rangle$ and which is coherently coupled to the Dicke state
$\ket{g'}$ by the dynamics. The control pulse $\Omega(t)$ shall then optimize
the dynamics by maximizing the fidelity
\begin{equation}
  \eta'=\sum_m|C_m|^2\eta_m^{(m)}\,,
\end{equation}
where $\eta_m^{(m)}=\bra{\psi^{(m)}}S(t)^\dagger|r_m\rangle\langle
r_m|S(t)\ket{\psi^{(m)}}$ and $S(t)$ is calculated for the effective
Hamiltonian of the atomic ensemble. The control field $\Omega(t)$ can be found
by means of an analogous strategy as for ensemble optimal control theory
  (OCT), finding the control pulse that optimizes the dynamics in each
subspace of $m$ excitations so to maximize $\eta'$
\cite{Rojan2014,Goerz2014,Kobzar2004,Kobzar2008,Koch2016}.

\section{Conclusions}
\label{Sec:Conclusions}
We have analysed the storage of a weak coherent pulse into the excitation of a
single atom inside a resonator, which acts as a quantum memory. Our specific
objective was to characterize the process in order to show under which
conditions an attenuated incident pulse can be considered as a single photon
for storage purposes. Thus we have identified the conditions and the figures of
merit which allow one to extract the single-photon storage fidelity by
measuring the probability that the atom has been excited at the end of the
process.

We remark that the retrieved information by a single atom will always be a
single photon~\cite{Chaneliere2005}. Nevertheless, the formalism we developed
in this work permits one to extend this dynamics to other kind of incident
pulses and to quantum memories composed by spin ensembles. For this general
case it sets the basis for identifying the optimal control pulses for storage
and retrieval of an arbitrary quantum light pulse.

\begin{acknowledgments}
  This work is dedicated to Wolfgang Schleich on the occasion of his 60th
  birthday. The authors are grateful to Stephan Ritter for insightful
  discussions and for proposing this problem. They also thank Susanne Blum,
  Peter-Maximilian Ney, Christiane Koch, and Gerhard Rempe for discussions. The
  authors acknowledge financial support by the German Ministry for Education
  and Research (BMBF) under the project Q.com-Q.
\end{acknowledgments}

\appendix
\section{\label{sec:descr-electr-field}Description of the electromagnetic field
  in the transmission line}
The transmission line is here modelled by a cavity of length $L$, with a
perfect mirror at $x=-L$ and the second mirror at $x=0$, which corresponds to
the optical cavity mirror with finite transmittivity. The modes of the
transmission line are standing waves with wave vector along the $x$ axis. For
numerical purposes we take a finite number $N$ of modes about the cavity wave
number $k_\mathrm{c}=\frac{\wc}{c}$. Their wave numbers are
\begin{equation}
  \label{eq:ffmodes}
  k_n =  k_\mathrm{c} + \frac{n\pi}{L}\,,
\end{equation}
and $n=-(N-1)/2,\dots,(N-1)/2$, the corresponding frequencies are
$\omega_n=ck_n$. We calibrate $N$ and $L$ so that our simulations are not
significantly affected by the finite size of the transmission line and by the
cutoff in the mode number $N$. For the propagation of the incident pulse and
its appropriate description at the mirror interface, this requires that the
difference between neighbouring frequencies is much smaller than the
characteristic frequencies of the problem. We further choose $N$ in order to
cover a frequency range which includes all the relevant frequencies of this
system. With the choice $|t_1|=t_2=6\Tc$, $L=12c\Tc$ and $N=311$, the norm of
the envelope results
\begin{equation}
  \label{eq:einnorm_approx}
  \int_{t_1}^{t_2}\abs{\Ein(t)}^2\de t = n(1-\varepsilon)
\end{equation}
with $\varepsilon<10^{-5}$. Further parameters and discussions are found in
Ref.~\cite{Giannelli2018}.

\section{Storage Efficiency for $n\ll1$. }
\label{App:B}

In this appendix we provide the details for calculating the dynamics and the
fidelity for an incident pulse which is a superposition of different photon
number states. We apply the procedure to multimode coherent states,
nevertheless it can be generalised in a straighforward manner to a generic
initial input pulse.

\subsubsection{\label{sec:decomp-coher-pulse}Decomposition of a coherent state}
The coherent state in Eq.~(\ref{eq:WCP_N_def}) can be decomposed in a linear
combination of states each with a fixed number of excitations (see
Eq.~(\ref{eq:WCP_N_deco_app}) with $C_m= \sqrt{e^{-n} n^{m}/m!}$): The mean
number of photons in the mode $k$ is $\abs{\alpha_k}^2$ and the mean photon
number in the coherent state is $n = \sum_{k=1}^N\abs{\alpha_k}^2$, see
Eq.~(\ref{eq:alpha}).
State $\ket{\psi^{(m)}}$ contains exactly $m$ excitations of the quantum
electromagnetic field and reads
\begin{subequations}
  \label{eq:psi_m}
  \begin{gather}
    \ket{\psi^{(0)}} {}=\ket{\text{vac}},\\
    \ket{\psi^{(1)}} {}=\sum_{k=1}^N\E_k\hat{b}_k^\dag\ket{\text{vac}},\\
    \ket{\psi^{(2)}} {}=\sum_{k=1}^N\sum_{k'=1}^N\E_{k,k'}\hat{b}_k^\dag
    \hat{b}_{k'}^\dag
    \ket{\text{vac}},\\
    \vdots \notag \\
    \ket{\psi^{(m)}} {}=\sum_{\{k\}_m}\E_{\{k\}_m}\overbrace{\hat{b}_k^\dag
      \hat{b}_{k'}^\dag \dots \hat{b}_{k'{'^\cdots}''}^\dag}^m\ket{\text{vac}}.
  \end{gather}
\end{subequations}
Coefficients $\E_{\{k\}_m}$ read
\begin{subequations}
  \label{eq:E_k_alpha_k}
  \begin{gather}
    \E_k = \frac{\alpha_k}{\sqrt{n}},  \label{eq:E_k_alpha_1}\\
    \E_{k,k'}=\E_{k',k}=\frac{\E_k\E_{k'}}{\sqrt{2}},\label{eq:E_k_alpha_2}\\
    \vdots  \\
    \E_{\{k\}_m}= \E_{\underbrace{_{k,k'\dots k'{'^\cdots}''}}_m} =
    \frac{\prod_{i\in \{k\}_m}\E_i}{\sqrt{m!}},
  \end{gather}
\end{subequations}
and it is easy to check that the states $\ket{\psi^{(m)}}$ are orthonormal
$\braket{\psi^{(m)}}{\psi^{(n)}}=\delta_{mn}$ and complete.

The storage fidelity when the initial state is the coherent state introduced in
Eq.~(\ref{eq:WCP_N_deco_app}) is given by (see
Eq.~(\ref{eq:fidelityexpansion}))
\begin{equation}
  \label{eq:fidelityexpansioncoherentpulse}
  \eta = e^{-n} \sum_{m=1}^\infty \frac{n^{m}}{m!}\eta^{(m)}.
\end{equation}

\subsubsection{\label{sec:decomposition-system}Equations of motion}

We here explicitly derive the equations of motion in the subspaces with
$m\leq2$ excitations.

\begin{figure}[!ht]
  \centering \includegraphics[]{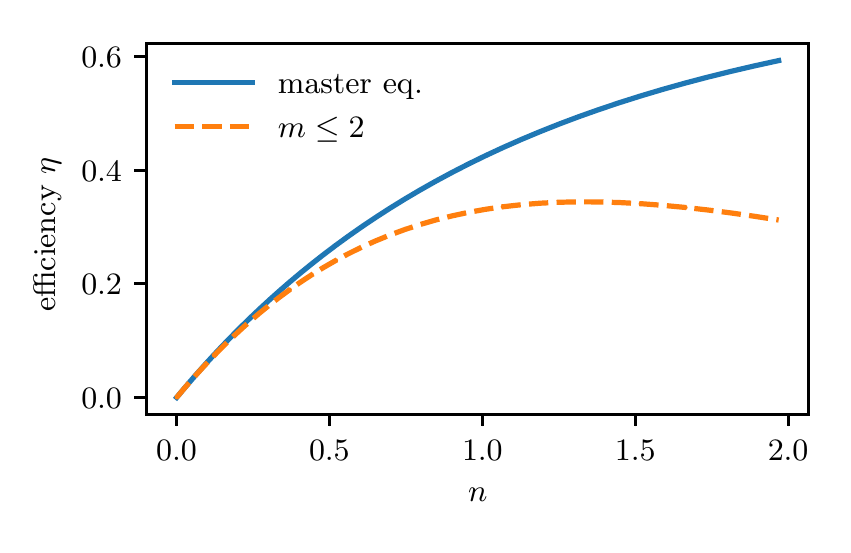}
  \caption{\label{fig:3}(Color online) Efficiency $\eta$ of the storage process
    of a weak coherent pulse. Solution with the master equation formalism of
    Sec.~\ref{sec:numerical-results} (solid line) and approximated solution
    with truncation to two excitations $m\leq2$ as described in the current
    section (dashed).}
\end{figure}

\textit{Zero excitations - Vacuum:} The subspace of zero excitations $m=0$
contains only the state $\ket{g,0,\text{vac}}$, meaning that the atom is in the
ground state $\ket{g}$, the cavity is empty and the electromagnetic field is in
the vacuum state. Thus the time evolution in this subspace is
$\ket{\phi^{(0)}_t}=\ket{\phi^{(0)}_{t_1}}$.

\textit{One excitation - Single photon:} A basis for the subspace with one
excitation $m=1$ is
\begin{equation*}
  \label{eq:basis1}
  \begin{aligned}
    \mathcal{B}_1=\{&\ket{g,1,\text{vac}},\ket{e,0,\text{vac}},
    \ket{r,0,\text{vac}},\\
    &\ket{g,0,1_k}: k\in \{1,\dots,N\}\}
  \end{aligned}
\end{equation*}
and a general state can be written as
\begin{equation}
  \label{eq:phi1}
  \begin{aligned}    
    \ket{\phi^{(1)}_t} =
    {}&c_1(t)\ket{g,1,\text{vac}}+e_1(t)\ket{e,0,\text{vac}}
    +{}\\
    &{}+r_1(t)\ket{r,0,\text{vac}}+\sum_k\mathcal{E}_k(t)\ket{g,0,1_k}.
  \end{aligned}
\end{equation}
The equations of motion in this subspace are ($\lambda_k=\lambda$)
\begin{equation}
  \label{eq:phi1_eqmotion_wdecay}
  \begin{gathered}
    \dot{c}_1(t) = -\I g e_1(t) -\I\lambda\sum_k\E_k(t)-\kbad c_1(t),\\
    \dot{e}_1(t) = (\I\Delta-\gamma)e_1(t) -\I g c_1(t) -\I\Omega(t)r_1(t),\\
    \dot{r}_1(t) = -\I\Omega^*(t)e_1(t),\\
    \dot{\E}_k(t) = -\I\Delta_k\E_k(t) -\I\lambda c_1(t),
  \end{gathered}
\end{equation}
and they constitute a system of $(N+3)$ coupled differential equations with
time dependent coefficients. Using the input output formalism~\cite{Walls1994}
one obtains
\begin{equation}
  \label{eq:phi1_inout}
  \begin{gathered}
    \dot{c}_1(t) =-\I g e_1(t)-\I\sqrt{2\kappa}\Ein(t)-(\kappa+\kbad )c_1(t),\\
    \dot{e}_1(t) =(\I\Delta -\gamma )e_1(t)-\I g c_1(t) -\I\Omega(t)r_1(t),\\
    \dot{r}_1(t) = -\I\Omega^*(t)e_1(t),
  \end{gathered}
\end{equation}
where $\kappa=L\lambda^2/c$ is the decay rate of the cavity field and $\Ein(t)$
is defined in Eq.~(\ref{eq:Einvsalphas}).
Equations~(\ref{eq:phi1_eqmotion_wdecay}) or Eqs.~(\ref{eq:phi1_inout}) can be
easily solved numerically. These equations correspond to the storage of a
single photon into a single atom~\cite{Giannelli2018} and are equivalent to the
approximated equations obtained in Ref.~\cite{Gorshkov2007a} describing the
storage of a light pulse in an atomic ensemble composed by a large number
$N\gg1$ of atoms.

\textit{Two excitations - Two photons states:} A basis for the subspace with
two excitations $m=2$ is
\begin{equation*}
  \begin{aligned}
    \mathcal{B}_2= \{&\ket{g,2,\text{vac}},\ket{g,1,1_k},
    \ket{g,0,1_k1_{k'}},\ket{e,1,\text{vac}}, \\
    &\ket{e,0,1_k},\ket{r,1,\text{vac}},\ket{r,0,1_k}: k,k'\in \{1,\dots,N\}\}
  \end{aligned}
\end{equation*}
thus a general state in this subspace can be written as
\begin{equation}
  \label{eq:phi2}
  \begin{aligned}
    \ket{\phi^{(2)}_t}={}&c_2(t)\ket{g,2,\text{vac}}+\sum_k\E_k^c(t)\ket{g,1,1_k}+{} \\
    &{}+\sum_{k}\sum_{k'\ge k}\E_{k,k'}(t)\ket{g,0,1_k1_{k'}} +{}\\
    &{}+ e_2(t)\ket{e,1,\text{vac}}+\sum_k\E_k^e(t)\ket{e,0,1_k} +{}\\
    &{}+ r_2(t)\ket{r,1,\text{vac}}+\sum_k\E_k^r(t)\ket{r,0,1_k}.
  \end{aligned}
\end{equation}
The state in Eq.~(\ref{eq:phi2}) can be used to describe the interaction of the
atom-cavity system with a two-photon state; in fact the term
$\sum_{k,k'}\E_{k,k'}(t)\ket{g,0,1_k1_{k'}}$ describes a two-photon state of
the electromagnetic field. Notice that we use the definition
$\ket{\cdot,\cdot,1_k1_{k'}} = b_k^\dag
b_{k'}^\dag\ket{\cdot,\cdot,\text{vac}}$ which implies
$\ket{\cdot,\cdot,1_k1_{k}} = \sqrt{2}\ket{\cdot,\cdot,2_k}$. The equations of
motion in this subspace are
\begin{equation}
  \label{eq:phi2_eqmotion_wdecay}
  \begin{aligned}
    &\begin{aligned}
      \dot{c_2}(t)=&-\I\sqrt{2} g e_2(t) -\I\sqrt{2}\lambda\sum_k\E_k^c(t)+{}\\
      &{}-2\kbad c_2(t)
    \end{aligned}\\
    &\begin{aligned}
      \dot{e_2}(t)={}&\br{\I\Delta-\gamma-\kbad } e_2(t) -\I \sqrt{2}g c_2(t)+{}\\
      &{}-\I\Omega(t)r_2(t) -\I\lambda\sum_k\E_k^e(t)
    \end{aligned} \\
    &\dot{r_2}(t) = -\I\Omega^*(t)e_2(t) -\I\lambda\sum_k\E_k^r(t)-\kbad r_2(t)\\
    &\begin{aligned}
      \dot{\E}_k^c(t) = &-\br{\I\Delta_k+\kbad }\E_k^c(t) -\I g\E_k^e(t)+{}\\
      &{}-\I\lambda\sum_{k'}A_{k,k'}(t)-\I\sqrt{2} \lambda c_2(t)
    \end{aligned}\\
    &\begin{aligned}
      \dot{\E}_k^e(t) ={}&\I\br{\Delta-\Delta_k}\E_k^e(t) -\I g\E_k^c(t)+{} \\
      &{}-\I\Omega(t)\E_k^r(t)-\I \lambda e_2(t)
    \end{aligned}\\
    &\dot{\E}_k^r(t) = -\I\Delta_k\E_k^r(t)-\I\Omega^*(t)\E_k^e(t)-\I \lambda r_2(t)\\
    &\begin{aligned}
      \dot{A}_{k,k'}(t) = {}&-\I\br{\Delta_k+\Delta_{k'}}A_{k,k'}+{}\\
      &{}-\I\lambda\br{\E_{k}^c(t)+\E_{k'}^c(t)},
    \end{aligned}
  \end{aligned}
\end{equation}
where we have defined $A_{k,k'}(t)=\E_{k,k'}(t)+\E_{k',k}(t)$.
Eqs.~(\ref{eq:phi2_eqmotion_wdecay}) are a system of $(N^2+3N+3)$ coupled
differential equations with time dependent coefficients; this system can be
solved numerically.

\textit{Calculation of the efficiency} The efficiency $\eta$ can be calculated
with the formalism introduced in this section in two ways: (i) solve
Eqs.~(\ref{eq:phi1_eqmotion_wdecay}) and Eqs.~(\ref{eq:phi2_eqmotion_wdecay})
with initial conditions given by the expansion~(\ref{eq:WCP_N_deco_app}) and
the coefficients given by Eqs.~(\ref{eq:E_k_alpha_1})
and~(\ref{eq:E_k_alpha_2}), then the efficiency is
\begin{equation}
  \label{eq:etacalc_1}
  \eta=\abs{r_1(t_2)}^2 +\abs{r_2(t_2)}^2 +\sum_k\abs{\E_k^r(t_2)}^2;
\end{equation}
or (ii) solve Eqs.~(\ref{eq:phi1_eqmotion_wdecay}) and
Eqs.~(\ref{eq:phi2_eqmotion_wdecay}) with initial
conditions~(\ref{eq:E_k_alpha_1}) and~(\ref{eq:E_k_alpha_2}) separately to
obtain the efficiencies $\eta^{(1)}$ and $\eta^{(2)}$ of single and double
photon storage; then the efficiency as function of $n$ is given by
Eq.~(\ref{eq:pop2vsalpha}).

Fig.~\ref{fig:3} reports the efficiency $\eta$ as a function of $n$, the solid
line represent the result of the numerical integration of the master equation
described in Sec.~\ref{sec:numerical-results}. The dashed line is the solution
with the decomposition until $m=2$ described in this section. It is evident
that for $n\ll1$ the two results coincide.


\end{document}